\documentstyle[11pt]{article}

\def\clsubset{\mathop{\kern0pt\subset}\limits _{\rm closed}  }

\newcounter{fig}
\newenvironment{numeriere}{\begin{list}{(\roman{fig})}{\usecounter{fig}
   \topsep-2ex \labelwidth 1cm\labelsep 0.5cm\leftmargin1.8cm}}{\end{list}}
\newcounter{satz}[section]
\newenvironment{Satz}[1]{\sl \refstepcounter{satz}\vskip1ex
{\bf \arabic{section}.\arabic{satz} #1. }}{\vskip1ex}
\newenvironment{Beweis}{\sl\vskip1ex {\bf Proof. }}{$\Box$\vskip1ex}

\newcommand{\NN}{I\!\!N}

\newcommand{\CC}{C\!\!\!\!I}

\newcommand{\A}{{\cal A}}     
\newcommand{\Ac}{{\bf A}}     

\newcommand{\End}{{\rm End}}
\newcommand{\Bb}{{\rm B}}
\newcommand{\hk}[1]{^{(#1)}}
\newcommand{\Le}{{\bf L}}
\newcommand{\oth}{\hat\otimes}

\def\Mittefrei#1#2{\hbox to \hsize{#1\hss#2}}

\begin{document}

\Mittefrei{\today}{MPI-Ph/93-63}
\Mittefrei{}{LMU-TPW 1993-23}
\vspace{15pt}

{\huge\sc
\renewcommand{\thefootnote}{\fnsymbol{footnote}}
\centerline{Unitary Continuous Representations}
\centerline{of Compact Quantum Groups}}\vspace{20pt}

{\Large\it
\begin{center}
Bernhard~Drabant {\large\it and} Wolfgang~Weich
\end{center} }
{\small\it
\begin{center}
Max-Planck-Institut f\"ur Physik,
F\"ohringer Ring 6, D-80805 M\"unchen\\
Sektion Physik der Universit\"at M\"unchen, Theresienstra\ss e 37,
D-80333 M\"unchen
\end{center} }
\vspace{25pt}

\noindent {\bf Abstract. }
{\rm
Generalizing the notion of continuous Hilbert space
representations of compact topological groups we define
unitary continuous correpresentations of $C^*$-completions of
compact quantum group Hopf algebras on arbitrary
Hilbert spaces.
It is proved that the unitary continuous correpresentations
decompose in finite dimensional irreducible correpresentations.
}\\[15pt]

\section{Introduction}
Continuous Symmetries play a fundamental r\^ole in physics.
In quantum mechanics the symmetries are represented
by unitary operators on the Hilbert space of states which is
in general infinite dimensional. This is the physical motivation
to study the unitary continuous representations of topological
groups. With the quantum groups a more general concept of symmetry
has been introduced into physics \cite{Drin,FRT,Woro}.
They are not any more
sets of transformations but are described by non-commutative
non-cocommutative $*$-Hopf algebras which should be understood in some
sense as the polynomials over the fictive quantum group.
Unitary finite dimensional correpresentations of
quantum groups have been studied in several papers,
e.~g. \cite{Koor,Rosso,Woro2}. The $*$-Hopf algebras with  $C^*$-norms
can be completed to $C^*$-algebras thus describing non-commutative
topological spaces \cite{Woro} ---
in the case of commutative $*$-Hopf algebras the
$C^*$-completions are the algebras of continuous functions over compact
topological groups.
For an application in quantum mechanics one is now interested in the
unitary continuous correpresentations on arbitrary Hilbert spaces
of those $C^*$-algebras arising from non-commutative non-cocommutative
$*$-Hopf algebras.

In section 2 correpresentations of $*$-Hopf algebras and the theory
of $*$-Hopf algebras of compact quantum groups
are reviewed \cite{Koor,Woro}.
With the help of the Haar functional
on a $*$-Hopf algebra of a compact quantum group the Hilbert space
completion is introduced in
section 3. The left action
by multiplication yields a $C^*$-norm.
These concepts are used in section 4
to define the continuous correpresentations of the $C^*$-completion.
In the concluding theorem it is stated that the unitary continuous
correpresentations on any Hilbert space decomposes into finite dimensional
correpresentations of the underlying $*$-Hopf algebra. This is a
generalization of the classical results on unitary representations of
compact topological groups \cite{Knapp}.

\section{The compact quantum group algebra $\A$}\label{A}
In this section definitions and results of \cite{Koor,Woro}
are quoted. Henceforth $\A$ is a $*$-Hopf algebra with comultiplication
$\Delta:\A\to\A\otimes\A$, counit $\epsilon:\A\to\CC$ and
antipode $S:\A\to\A$.

For some vector space $V$ let $\End(V)$ denote its endomorphisms.
The transposition operator $\sigma_V$ acts as endomorphism on
$V\otimes V$ by $a\otimes b\mapsto b\otimes a$.

\begin{Satz}{Definition}
A correpresentation $(T,V)$ of $\A$ on the vector space $V$
is given by
\begin{numeriere}
\item $T\in\A\otimes \End(V)$, such that
\item $T_{13} T_{23} = (\Delta\otimes id) (T)$,
 $T_{23} := 1_\A\otimes T$, $T_{13}:= (\sigma_\A\otimes id)(T_{23})$,
\item $(\epsilon\otimes id) (T) = 1_{\End(V)}$.
\end{numeriere}
If $V$ is a Hilbert space
the correpresentation $(T,V)$ of $\A$ is called unitary if
$T^* T = 1_\A\otimes 1_{\End(V)} = T\, T^*$,
where $T^*:= (^*\otimes^*)(T)$.
\end{Satz}

If $dim(V) =n<\infty$ one can choose a basis $\{e_i\}_{i=1,..,n}$ and
a dual basis $\{e^*_i\}_{i=1,..,n}$ with $e^*_i(e_j) = \delta_{ij}$.
One can expand $T=\sum_{i,j=1}^n t_{ij}\otimes (e_i\otimes e^*_j)$ and
obtains in particular
$\Delta(t_{ij}) = \sum_k t_{ik}\otimes t_{kj}$ and
$\epsilon(t_{ij})=\delta_{ij}$. If $V$ is a Hilbert space, $(T,V)$ a unitary
correpresentation, then $e^*_i = (e_i|.)$ and $S(t_{ij})=t^*_{ji}$.

\begin{Satz}{Definition} \label{Def-iv}
Let $(T,V)$ be a correpresentation of $\A$ on the
vector space $V$. A linear subspace $W$ of $V$ is
called invariant if $T(1_\A\otimes W)\subset\A\otimes W$. \\
$(T,V)$ is called irreducible if $V$ and $\{0\}$ are the only invariant
subspaces of $V$. \\
Let $(S,W)$ be another correpresentation of $\A$. A linear operator
$L: V\to W$ is called an intertwiner if
$ S (1_\A\otimes L) = (1_\A\otimes L) T$. \\
$(T,V)$ and $(S,W)$ are called equivalent if there exists a bijective
intertwiner.
\end{Satz}

\begin{Satz}{Schur's Lemma}
Let $L: V\to W$ be an intertwiner for the correpresentations
$(T,V)$ and $(S,W)$ of $\A$. If both are irreducible
then either $L=0$ or $L$ is bijective.
\end{Satz}

\begin{Satz}{Definition}
The collection of all equivalence classes of irreducible unitary finite
dimensional correpresentations of $\A$ is called $\hat\A$.
\end{Satz}

Any $*$-Hopf algebra has in particular the scalar unitary correpresentation
$(1_\A,\CC)$. We denote its equivalence class by $\hat 0\in\hat\A$.
For each $\alpha\in\hat\A$ we choose a representative $(T^\alpha,V^\alpha)$
which is unitary.

\begin{Satz}{Proposition}
For each $\alpha\in\hat\A$ let
$\{t^\alpha_{ij}\}_{i,j}$ be the set of
matrix elements of $(T^\alpha,V^\alpha)$.
Then $\{t^\alpha_{ij}\}_{\alpha\in\hat\A,i,j}$ is a linear independent set.
\end{Satz}

Setting
$\A^\alpha:=span \{t^\alpha_{ij}\}_{ij}\subset \A$
for $\alpha\in \hat\A$
the following definition is motivated by the
theory of compact topological groups.

\begin{Satz}{Definition}
A $*$-Hopf algebra $\A$
is a compact quantum group algebra if it is the linear
span of the matrix elements of its unitary
irreducible finite-dimensional correpresentations:
${\A=\bigoplus_{\alpha\in\hat\A} \A^\alpha}$.
\end{Satz}

The algebra of characteristic functions on a compact
topological group is such a compact quantum group algebra.
The Hopf algebras of the
quantum groups $SU_q(N)$ and $SO_q(N)$ ($q>0$) of \cite{FRT} are
non-commutative examples.

Having a decomposition of $\A$ the Haar functional can be defined.

\begin{Satz}{Definition}
The linear functional $h: \A\to\CC$ with $h(1_\A)=1$ and
$h_{|\A^\alpha}=0$
for $\alpha\ne \hat 0$ is called the Haar functional.
\end{Satz}

\begin{Satz}{Theorem}
The Haar functional $h$ on the compact quantum group algebra $\A$
is uniquely determined by its properties
\begin{equation}
 (h\otimes id)(\Delta(a))=h(a) \,1_\A=(id\otimes h)(\Delta(a))
 \quad\mbox{and}  \quad
 h(1_\A)=1.
\end{equation}
The Haar functional is positive: $h(a^* a) > 0$ if $a\ne 0$.
\end{Satz}

\section{The completions of $\A$}\label{B}
In sections 3. and 4. $\A$ is always understood to be a compact quantum group
algebra with Haar functional $h$. This Haar functional
defines an inner product
$\A\times \A \ni (a,b) \mapsto (a|b) := h(a^* b)$
which makes $\A$ a pre-Hilbert space. Thus also $\A\otimes...\otimes \A$
becomes naturally a pre-Hilbert space.
The Hilbert space completion of the pre-Hilbert spaces $X$ and $Y$ is
denoted by $X\oth Y$,
the Hilbert space completion of the direct sum
of the pre-Hilbert spaces $X_i,\;i\in I$ is denoted by
$\hat\oplus_{i\in I} X_i$. For a normed vector space $V$ we denote
its continuous endomorphisms with $\Bb(V)$.

\begin{Satz}{Definition}
For each $n\in\NN$ let
$\ell^{\oth n}$ be the Hilbert space completion of the $n$-fold
tensor product of $\A$, let $\ell^{\oth 0}:=\CC$.
\begin{equation}
 \Le := \bigoplus^\wedge_{n\in\NN_0} \ell^{\oth n}.
\end{equation}
\end{Satz}

$\ell:=\ell^{\oth 1}$ corresponds to the square
integrable functions over the quantum group.

$\A$ acts on itself by left multiplication. This yields a
$*$-representation $\A\ni a\mapsto \lambda(a)\in \Bb(\A)$,
$\lambda(a): b\mapsto ab\in\A$.
Because of the unitarity condition
$\sum_i t^{\alpha\,*}_{ij}t^\alpha_{ik} = \delta_{jk}$
in $\A^\alpha$ one gets that $\lambda(a)$
is a bounded operator on the pre-Hilbert space $\A$
and can therefore be extended
to $\ell$, i.~e.
$\A\ni a\mapsto \lambda(a)\in \Bb(\ell)$ is a faithful Hilbert space
representation of $\A$ \cite{Koor}.
Counit and comultiplication are $*$-algebra
homomorphisms and serve to define Hilbert space representations of
$\A$ on $\ell^{\oth n}$ and $\Le$.

\begin{Satz}{Definition}
$\Delta\hk 0:=\epsilon$. For $n>0$, $\Delta\hk n:=
(\Delta\hk{n-1}\otimes id)\circ \Delta$.
$\Lambda\hk n:=\lambda^{\otimes n}\circ\Delta\hk n:\A\to\Bb(\ell^{\oth n})$.
$\Lambda := \bigoplus_{n=0}^\infty \Lambda\hk n: \A\to \Bb(\Le)$ is the
direct sum Hilbert space representation of $\A$ on $\Le$.
\end{Satz}

This justifies the following definition.

\begin{Satz}{Definition}
The operator norm $|a| := \sup_{x\in\Le,||x||_\Le=1} ||\Lambda(a) x||_\Le
= ||\Lambda(a)||_\Le$ is
a $C^*$-norm on $\A$. The corresponding completion of $\A$ is the
$C^*$-algebra $\Ac$. $\Lambda(\Ac)$ and $\Ac$ are identified.
\end{Satz}

\begin{Satz}{Proposition}
Comultiplication and counit on $\A$ are bounded and can be
extended to continuous mappings on $\Ac$,
$\Delta:\Ac\to\overline{\Ac\otimes\Ac}\clsubset \Bb(\Le\otimes\Le)$,
$\epsilon: \Ac\to\CC$.
\end{Satz}

\begin{Beweis}
The boundedness of $\epsilon$ follows from $|a|=||\Lambda(a)||_{\Le}\ge
||\Lambda(a)||_{\ell^{\oth 0}} = |\epsilon(a)|$. The boundedness of $\Delta$
comes from $||(\Lambda\otimes\Lambda)(\Delta(a))||_{\Le\oth\Le} =
||\Lambda(a)||_{\Le}$.
\end{Beweis}

The antipode need not be continuous and can therefore not be extended on $\Ac$.
If $\A$ is the linear span of the matrix elements of the unitary irreducible
group representations of a compact topological group then
$\Ac$ is the $C^*$-algebra of continuous functions over this group. This is
the contents of the Peter-Weyl Theorem, see e.~g. \cite{Knapp}.

With the above construction the compact quantum group algebra $\A$
determines uniquely the $C^*$-completion $\Ac$.

\begin{Satz}{Lemma}\label{h-extension}
The Haar functional $h$ is bounded on $\A$ respective to the norms
$||.||_\ell$ and $|.|$. Its extension on $\ell$ is given by
$\ell\ni x\mapsto (1_\A| x)$.
Denoting the extension of $h$ on the
$C^*$-algebra by $h: \Ac\to\CC$ the following holds
\begin{equation}\label{h-ext-form}
 \forall a\in\Ac:\, h(a) = (1_\A|\Lambda(a) 1_\A) .
\end{equation}
\end{Satz}

\section{Continuous correpresentations} \label{C}
Now we can define continuous correpresentations of $\Ac$. Since
$\sigma_{\Bb(L)}$ is bounded on $\Bb(\Le)\otimes \Bb(\Le)\subset
\Bb(\Le\oth\Le)$
it can be continued on the closure.

\begin{Satz}{Definition}
A continuous correpresentation $(U,H)$ of $\Ac$ on the Hilbert space
$H$ is given by
\begin{numeriere}
\item $U\in\overline{\Ac\otimes \Bb(H)}\clsubset\Bb(\Le\oth H)$, such that
\item $U_{13} U_{23} = (\Delta\otimes id) (U)$,
 $U_{23} := 1_\Ac\otimes U$, $U_{13}:= (\sigma_\Ac\otimes id)(U_{23})$,
\item $(\epsilon\otimes id) (U) = 1_{\Bb(H)}$.
\end{numeriere}
Let $U^*$ denote the adjoint of $U$ in $\Bb(\Le\oth H)$. \\
The continuous correpresentation $(U,H)$ of $\Ac$ is called isometric if
$U^* U = 1_\Ac\otimes 1_{\Bb(H)}$. \\
$(U,H)$ is called unitary if $U\, U^* = 1_\Ac\otimes 1_{\Bb(H)} = U^* U$.
\end{Satz}

In the case of a commutative $*$-Hopf algebra $\A$ standing for the
polynomials over a compact topological group $G$ this definition
is equivalent to the statement that
the mapping $G\times H\to H$, $(g,v)\mapsto U(g)(v)$ is a continuous
representation of $G$ on $H$.

If $H$ is finite dimensional then $U\in\Ac\otimes \Bb(H)$.
It is evident that every correpresentation of $\A$ on a Hilbert space
is in particular a continuous correpresentation of $\Ac$.
The definition 2.\ref{Def-iv} of
invariant subspaces and irreducibility
can be taken over to the case of $\Ac$.

\begin{Satz}{Definition}
Let $(U,H)$ be a continuous correpresentation of $\Ac$.
The closed subspace $K\clsubset H$ is called invariant if the orthogonal
projector $P_K\in \Bb(H)$, $P_K H = K$ fulfills
\begin{equation}
 U (1_\Ac\otimes P_K) =  (1_\Ac\otimes P_K)  U (1_\Ac\otimes P_K) =: U_K.
\end{equation}
In this case $(U_K,K)$ is called a continuous subcorrepresentation
of $\Ac$. \\
The continuous correpresentation $(U,H)$ is called irreducible if the only
invariant subspaces of $H$ are $H$ and $\{0\}$. \\
Let $(U,H)$ and $(Q,N)$ be continuous correpresentations of $\Ac$ on
the Hilbert spaces $H$ and $N$, let $L: H\to N$ be linear and continuous.
If
\begin{equation} \label{intertwiner}
 Q (1_\Ac\otimes L) = (1_\Ac\otimes L) U
\end{equation}
then $L$ is called an intertwiner.
\end{Satz}

These definitions are justified by the following observation which can
easily be verified.

\begin{Satz}{Lemma} \label{subcorrepresentation}
The continuous subcorrepresentation $(U_K,K)$ is a
continuous correpresentation.
If $(U,H)$ is isometric then each continuous subcorrepresentation is
also isometric.
\end{Satz}

\begin{Satz}{Schur's Lemma}
Let $L: H\to N$ be an intertwiner for the correpresentations
$(U,H)$ and $(Q,N)$ of $\Ac$ on the Hilbert spaces
$H$ and $N$ with closed image $L(H)\clsubset N$.
If both are irreducible then either $L=0$ or $L$ is bijective.
\end{Satz}

\begin{Beweis}
Define in $\Bb(H)$ the orthogonal projector $P_0$, $P_0 H=ker L$ and in
$\Bb(K)$ the orthogonal projector $P_1$, $P_1 N=L(H)$. Apply to
equation (\ref{intertwiner}) $1_\Ac\otimes P_0$ from the right or
$1_\Ac\otimes P_1$ from the left respectively to prove that $ker L$ and
$L(H)$ are invariant. If $L\ne 0$ then $ker L = \{0\}$ because of the
irreducibility of $(U,H)$. If $H\ne\{0\}$ it then follows from the
irreducibility of $(Q,N)$ that $N=L(H)$.
\end{Beweis}

\begin{Satz}{Lemma} \label{U-finite}
Let $H$ be a finite-dimensional Hilbert space, $(U,H)$
an irreducible isometric continuous correpresentation of $\Ac$.
\begin{numeriere}
\item $U\in\A\otimes \Bb(H)$, and $(U,H)$ is
an irreducible correpresentation of $\A$.
\item $(U,H)$ is unitary, $(S\otimes id) (U) = U^*$.
\end{numeriere}
\end{Satz}

\begin{Beweis}
Let $\{e_A\}_{A=1,..,d}$ be an orthonormal basis in $H$, and $e_A^*:=
(e_A|.)$ the dual basis.
Expand $U=\sum_{A,B=1}^d u_{AB}\otimes (e_A\otimes e_B^*)\in \Ac\otimes
\Bb(H)$.
Consider the following matrices
$M\hk{\alpha,j,B}_{iA} := h\left(t^{\alpha \,*}_{ji} u_{BA}\right)$.
There is at least one $\alpha_u\in\hat\A$
and $j,B$ such that $M\hk{\alpha_u,j,B} \ne 0$.
The opposite would imply the contradiction
\begin{eqnarray}
 0 &=& \sum_B \sum_{\alpha,i,j} h\big(u^*_{BA} t^{\alpha}_{ij}\big) \,
 \frac{1}{||t^\alpha_{ij}||^{2}_\ell} h\big(t^{\alpha\,*}_{ij} u_{BA}\big)
 \nonumber \cr
   &=& \sum_B \sum_{\alpha,i,j}
  \big(\Lambda(u_{BA}) 1_\A\big|t^{\alpha}_{ij}\big)
 \frac{1}{||t^\alpha_{ij}||^{2}_\ell}
  \big(t^{\alpha}_{ij}\big|\Lambda(u_{BA})1_\A\big)
  \cr
  &=& \sum_B \big(\Lambda(u_{BA}) 1_A\big|\Lambda(u_{BA}) 1_A\big) =
     \sum_B \big(1_\A\big|\Lambda(u^*_{BA} u_{BA}) 1_\A\big) = 1
  \nonumber
\end{eqnarray}
where Lemma \ref{B}.\ref{h-extension} and the isometry of $(U,H)$
have been used. In $V^{\alpha_u}$
an orthonormal basis $\{f_i\}$ with $f_i^* := (f_i|.)$ such that
$T^{\alpha_u} = \sum_{ij} t^\alpha_{ij}\otimes (f_i\otimes f_j^*)$ is used.
Then $L := \sum_{iA} M\hk{\alpha_u,j,B}_{iA} (f_i\otimes e_A^*)$ is a
non-vanishing
intertwiner $(1_\Ac\otimes L) U = T^{\alpha_u} (1_\Ac\otimes L)$.
Since $L(H)$ is closed Schur's Lemma implies that
$L$ is invertible, and one gets
$U=(1_\A\otimes L^{-1}) T^{\alpha_u} (1_\A\otimes L) \in
\A\otimes \Bb(H)$ thus proving the first statement.
Applying the antipode to the last equation one gets
$(S\otimes id)(U) = (1_\A\otimes L^{-1}) T^{\alpha_u\,*}
(1_\A\otimes L) = U^*$ and hence
the unitarity of $U$.
\end{Beweis}

Statement (i) of Lemma 4.\ref{U-finite} implies that the $*$-Hopf algebra
can be uniquely reconstructed from its $C^*$-completion $\Ac$.

\begin{Satz}{Proposition}
Let $(U,H)$ be an isometric continuous
correpresentation on the Hilbert space $H$,
$K$ a finite dimensional invariant subspace such that
the continuous subcorrepresentation $(U_K,K)$ is irreducible.
\begin{numeriere}
\item
$(U_K,K)$ is a unitary continuous correpresentation of $\Ac$.
\item
$U (1_\Ac\otimes P_K) = (1_\Ac\otimes P_K) U$ .
\item
$H\ominus K$ is invariant,
$((1_\Ac\otimes (1_{\Bb(H)}-P_K))U,H\ominus K)$ is an isometric continuous
correpresentation of $\Ac$.
\item
If $H$ is finite dimensional then it decomposes in
irreducible continuous subcorrepresentations, and $U\in\A\otimes \Bb(H)$.
\end{numeriere}
\end{Satz}

\begin{Beweis}
The first statement follows from Lemmas 4.\ref{subcorrepresentation}
and 4.\ref{U-finite}. One gets therefore
$1_\Ac\otimes P_K = U_K\, U_K^* = U(1_\Ac\otimes P_K)U^*$.
Multiplying with $U$
from the right proves the second statement. The third one is a direct
consequence using Lemma 4.\ref{subcorrepresentation}.
The last assertion follows by induction.
\end{Beweis}

The last statement means that the finite dimensional unitary
continuous correpresentations of $\Ac$ coincide with the finite dimensional
unitary correpresentations of $\A$.

\begin{Satz}{Definition}
Let $f\in\A$, $(U,H)$ an isometric continuous correpresentation of $\Ac$
on the Hilbert space $H$.
\begin{equation}
 h_f := (h\otimes id) \left((f\otimes 1_{\Bb(H)}) U\right) \in \Bb(H).
\end{equation}
\end{Satz}

\begin{Satz}{Proposition} \label{hf}
Let $(U,H)$ be an isometric continuous correpresentation of $\Ac$
on the Hilbert space $H$. Let
$\left\{(U_{K_\rho},K_\rho)\right\}_{\rho\in R}$
be a family of irreducible finite dimensional continuous
subcorrepresentations, let
$P_\rho := P_{K_\rho}$ be
the orthogonal projectors on $K_\rho$ respectively.
\begin{numeriere}
\item The norm of $h_f$ is bounded by $||h_f||_H\le ||f||_\ell$.
\item $P_\rho h_f = h_f P_\rho. $
\item $U (1_\Ac\otimes h_f) = \sum_A S(f'_A) \otimes h_{f''_A}\,$ where
 $\Delta(f) = \sum_A f'_A \otimes f''_A$.
\item If $v\ne 0$ then there is an $f\in\A$ such that $h_f(v)\ne 0$.
\end{numeriere}
\end{Satz}

\begin{Beweis}
Apply $U$ to $1_\A\otimes v \in \ell\oth H \subset \Le\oth H$.
$U(1_\A\otimes v) =: \sum_i u_i\otimes v_i \in \ell\oth H$ such that
the $v_i\in H$ are mutually orthogonal. Then $h_f(v) = \sum_i (f^*|u_i) v_i$.
The first assertion follows
from $||h_f(v)||_H^2 = \sum_i |(f^*|u_i)|^2 ||v_i||_H^2 \le
\sum_i(f^*|f^*) (u_i|u_i)\,
||v_i||_H^2 = ||f^*||^2_\ell ||v||_H^2$. The second is obvious.
$U (1_\Ac\otimes h_f) = (id\otimes h\otimes id)
\left( (1_\Ac\otimes f\otimes 1_{\Bb(H)}) U_{13} U_{23} \right) =
(id\otimes h\otimes id)
\sum_A\left( (\epsilon(f'_A)\,1_\Ac\otimes f''_A\otimes 1_{\Bb(H)})
(\Delta\otimes id)(U) \right)$ proves the third statement.
$h_f(v)=0$ is equivalent to $(f^*|u_i)=0$ for all $i$. Since $\A$ is
dense in $\ell$ the forth statement follows.
\end{Beweis}

\begin{Satz}{Theorem}
Let $(U,H)$ be an isometric continuous correpresentation of $\Ac$ on the
Hilbert space $H$. Then $(U,H)$ is unitary and $H$ decomposes in finite
dimensional irreducible continuous subcorrepresentations
$(U_\rho,K_\rho)$, $U_\rho\in\A\otimes \Bb(K_\rho)$:
\begin{equation}
 H = \bigoplus^\wedge_{\rho\in R} K_\rho.
\end{equation}
\end{Satz}

\begin{Beweis}
Using Zorn's Lemma let $R$ be a maximal index set such that
$P_\rho$ with $\rho\in R$ are projectors on finite dimensional irreducible
subspaces of $H$ with $P_\rho P_{\rho'} = 0$ if $\rho\ne\rho'$.
Then choose $v\in (1_{\Bb(H)}-\sum_{\rho\in R} P_\rho) H$.
{}From proposition 4.\ref{hf} it follows that $h_f(v)$ is contained in a
finite dimensional continuous correpresentation. Hence $v$ is necessarily
$0$.
Since the restriction of $(U,H)$ to each irreducible continuous
subcorrepresentation
is unitary the full continuous correpresentation is necessarily unitary.
\end{Beweis}

The theory of unitary continuous correpresentations of the $C^*$-completions
of the $*$-Hopf algebras of compact quantum groups is thus reduced to the
correpresentation theory of the underlying $*$-Hopf algebras.
In \cite{Woro2} S.~L.~Woronowicz showed that the theory of finite dimensional
correpresentations of the $*$-Hopf algebra of $SU_q(N)$ coincides with the
theory for the undeformed $SU(N)$. Together with our results one concludes
therefore that also the theory of unitary continuous correpresentations of
these quantum groups coincides with the theory for the undeformed Lie groups.

The authors acknowledge gratefully discussions with M.~Dijkhuizen,
T.~Koornwinder and J.~Wess.

\pagebreak


\begin{thebibliography}{999}
\bibitem{Drin} V.~G.~Drinfeld, Proceedings of the International Congress of
Mathematicians,
Berkeley, California, 798 (1986)
\bibitem{FRT} L.~D.~Faddeev, N.~Yu.~Reshetikhin, L.~A.~Takhtajan,
{\it Algebra and Analysis} {\bf 1}, 178 (1989)
\bibitem{Knapp} A.~W.~Knapp, Lie groups, Lie algebras, and cohomology,
{\it Mathematical Notes} {\bf 34}, Princeton 1988
\bibitem{Koor} T.~H.~Koornwinder, Lectures at the European School of
Group Theory, Trento 1993\\
M.~S.~Dijkhuizen, Thesis, CWI Amsterdam 1993
\bibitem{Rosso} M.~Rosso, {\it Comm.~Math.~Phys. }{\bf 117}, 581 (1988)
\bibitem{Woro} S.~L.~Woronowicz, {\it Comm.~Math.~Phys.} {\bf 111}, 139 (1987)
\bibitem{Woro2} S.~L.~Woronowicz, {\it Invent.~math.} {\bf 93}, 35 (1988)
\end{thebibliography}
\end{document}